\documentclass[preprintnumbers,prl,twocolumn,showpacs,letterpaper,nofootinbib]{revtex4-1}

\usepackage[pdftex]{graphicx}
\usepackage{latexsym,amsmath,amssymb,lmodern,float,url}
\usepackage{natbib}
\usepackage[pdftex,bookmarks,linktocpage,pdfpagelabels,plainpages=false,hyperfigures,linkcolor=blue,citecolor=blue]{hyperref} 
\hypersetup{colorlinks=true}

 \usepackage{scrextend}
 \usepackage[percent]{overpic}

\newcommand{\mm}{(\mu^+\mu^-)}

\begin{document}
\title{Discovering True Muonium in $K_L\rightarrow \mm \gamma$}

\author{Yao Ji}
\email{yao.ji@physik.uni-regensburg.de}
\affiliation{Institut f\"ur Theoretische Physik, Universit\"at Regensburg, Regensburg 93040, Germany}
\author{Henry Lamm}
\email{hlamm@umd.edu}
\affiliation{Department of Physics, University of Maryland, College Park, MD 20742}
\date{\today}

\begin{abstract}
Theoretical and phenomenological predictions of $\mathcal{BR}(K_L\rightarrow \mm \gamma)\sim7\times10^{-13}$ are presented for different model form factors $F_{K_L\gamma\gamma^*}(Q^2)$.  These rates are comparable to existing and near-term rare $K_L$ decay searches at J-PARC and CERN, indicating a discovery of true muonium is possible.  The model uncertainties are sufficiently small that detection of true muonium could discriminate between the form factor models.  Further discussion of potential backgrounds is made.
\end{abstract}

\maketitle

Lepton universality predicts differences in electron and muon observables should occur only due to their mass difference.  Measurements of $(g-2)_\ell$~\cite{PhysRevD.73.072003}, nuclear charge radii~\cite{Antognini:1900ns,Pohl1:2016xoo}, and rare meson decays~\cite{Aaij:2014ora,*Aaij:2015yra} have shown hints of violations to this universality.  The bound state of $\mm$, \textit{true muonium}, presents a unique opportunity to study lepton universality in and beyond the Standard Model~\cite{TuckerSmith:2010ra,*Lamm:2015gka,*Lamm:2016jim}.  To facilitate these studies, efforts are on-going to improve theoretical predictions~\cite{Jentschura:1997tv,*Jentschura:1997ma,*PhysRevD.91.073008,*Lamm:2016vtf,*PhysRevA.94.032507,*Lamm:2017lib}.  Alas, true muonium remains undetected today.

Since the late 60's, two broad categories of $\mm$ production methods have been discussed: particle collisions (fixed-target and collider)~\cite{Bilenky:1969zd,*Hughes:1971,*Moffat:1975uw,*Holvik:1986ty,*Ginzburg:1998df,*ArteagaRomero:2000yh,*Brodsky:2009gx,*Chen:2012ci}, or through rare decays of mesons~\cite{Nemenov:1972ph,*Vysotsky:1979nv,*Kozlov:1987ey,Malenfant:1987tm}.  Until recently, none have been attempted due to the low production rate ($\propto \alpha^4$).  Currently, the Heavy Photon Search (HPS)~\cite{Celentano:2014wya} experiment is searching for true muonium~\cite{Banburski:2012tk} via $e^-Z\rightarrow\mm X$. Another fixed-target experiment, but with a proton beam, DImeson Relativistic Atom Complex (DIRAC)~\cite{Benelli:2012bw} studies the $(\pi^+\pi^-)$ bound state and could look for $\mm$ in a upgraded run~\cite{dirac}.  

In recent years, a strong focus on rare kaon decays has developed in the search for new physics.  The existing KOTO experiment at J-PARC~\cite{Ahn:2016kja} and proposed NA62-KLEVER at CERN~\cite{Moulson:2016zsl} hope to achieve sensitivities of $\mathcal{BR}\sim10^{-13}$ allowing a 1\% measurement of $\mathcal{BR}(K_L\rightarrow\pi^0\nu\nu)\sim10^{-11}$.  Malenfant was the first to propose $K_L$ as a source of $\mm$~\cite{Malenfant:1987tm}. He estimated $\mathcal{BR}(K_L\rightarrow\mm \gamma)\sim5\times10^{-13}$ by approximating $F_{K_L\gamma\gamma*}(Q^2=4M_\mu^2)\sim F_{K_L\gamma\gamma*}(0)$ where $Q^2$ is the off-shell photon invariant mass squared. This two-body decay is the reach of rare kaon decay searches and is an attractive process for discovering $\mm$.  The decay has simple kinematics with a single, monochromatic photon (of $E_\gamma=203.6$ MeV if the $K_L$ is at rest) plus $\mm$ which could undergo a two-body dissociate or decay into two electrons (with $M_{\ell\ell}^2\sim4M_\mu^2$). 

Another motivation for the search for this rare decay is its unique dependence on the form factor.  Previous extractions of the form factor relied upon radiative Dalitz decays, $K_L\rightarrow \ell^+\ell^-\gamma$, the most recent being from the KTEV collaboration~\cite{Abouzaid:2007cm,AlaviHarati:2001wd}.  In these analyses, the phenomenological form factor is integrated over bins in $Q^2$, and fit to differential cross section data.  In contrast, the $\mm$ branching ratio gives the form factor at one $Q^2$ and fixes one-parameter form factors.  Further, a measurement of $\mm$ would help to better understand the kaon form factor through a completely different set of systematic and statistical uncertainties to the existing measurements.

In this letter, we present the $\mathcal{BR}(K_L\rightarrow\mm \gamma)$ including full $\mathcal{O}(\alpha)$ radiative corrections and four different treatments of the form factor $F_{K_L\gamma\gamma*}(Q^2)$, thereby avoiding Malenfant's approximation.   It is shown that the approximation underestimates the branching ratio by a model-dependent 15-60\%.  Possible discovery channels are discussed and brief comments on important backgrounds are made.
 
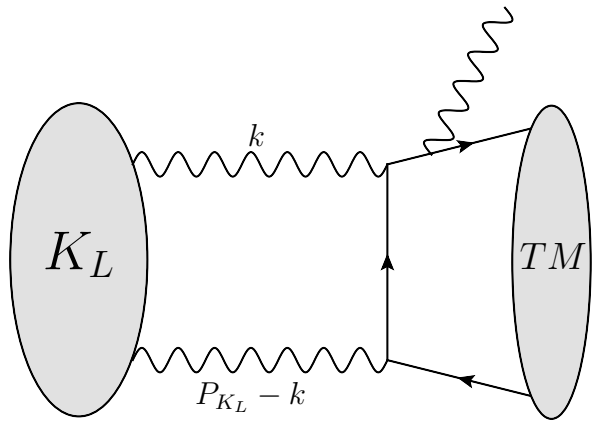
\begin{figure}[!t]
\begin{overpic}[width=.9\linewidth]{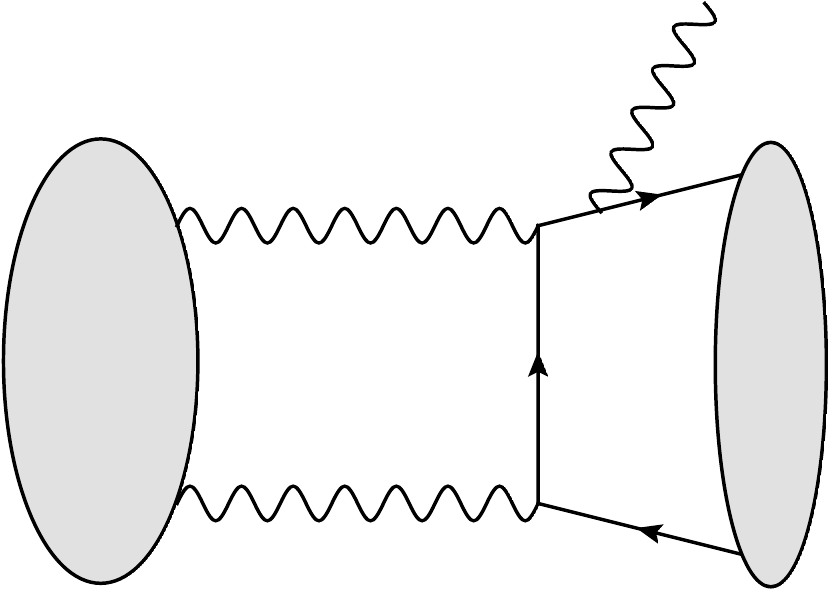}
 \put (6,26) {\huge$\displaystyle K_L$}
 \put (87.5,26) {\Large$\displaystyle TM$}
\put (41,47) {\large$\displaystyle k$}
\put (32,3) {\large$\displaystyle P_{K_L}-k$}
 \end{overpic}
\caption{Feynman diagram of $K_L\rightarrow\gamma^*\gamma^*\rightarrow\mm\gamma$ which contributes to the branching ratio at ${\cal O}(\alpha^5)$ and is proportional to $F_{\gamma^*\gamma^*}(z_1,z_2)$}
\label{fig:Oalpha-additonal}
\end{figure}

Following previous calculations for atomic decays of mesons~\cite{Nemenov:1972ph,Vysotsky:1979nv,Kozlov:1987ey,Malenfant:1987tm}, the branching ratio can be computed
\begin{align}
\label{eq:br1}
\frac{\mathcal{BR}(K_L\rightarrow\mm\gamma)}{\mathcal{BR}(K_L\rightarrow\gamma\gamma)}=&\nonumber\\\frac{\alpha^4\zeta(3)}{2}\left(1-z_{TM}\right)^3\bigg[1&-\frac{0.439\alpha}{\pi}\bigg]|f\left(z_{TM}\right)|^2\, ,
\end{align}
where $\zeta(3)=\sum_n 1/n^3$ arising from the sum over all allowed $\mm$ states, $z_{TM}=M_{TM}^2/M_K^2\approx4M_{\mu}^2/M^2_K$, and $f(z)=F_{K_L\gamma\gamma^*}(z)/F_{K_L\gamma\gamma^*}(0)$.  Previous computation of radiative corrections considered only the vacuum polarization from the flavor found in the final state~\cite{Vysotsky:1979nv}.  We have computed the full $\mm$ results including the electronic, muonic, and hadronic vacuum polarization~\cite{PhysRevA.94.032507} as well as the QED process $K_L \rightarrow \gamma^*(k)+\gamma^*(P_{K_L}-k)\rightarrow\gamma + \text{TM}$ demonstrated by Fig.~\ref{fig:Oalpha-additonal} where $P_{K_L}$ is the four-momentum of the $K_L$. In this contribution, one should take the convolution of the QED amplitude with double-virtual-photon form factor $F_{K_L\gamma^*\gamma^*}(k^2/M_K^2,(P_I-k)^2/M_K^2)$. For our purpose, however, taking the form factor to be $F_{\gamma\gamma^*}(0,z_{TM})$ is a sufficient approximation as shown in~\cite{Kampf:2005tz}.  A similar calculation for positronium, where other lepton flavors and hadronic loop corrections are negligible, finds the $\displaystyle\frac{\alpha}{\pi}$ coefficient is $-52/9$.  

$F_{K_L\gamma\gamma^*}(0)$ is fixed to the experimental value of $\mathcal{BR}(K_L\rightarrow\gamma\gamma)=5.47(4)\times10^{-4}$~\cite{Olive:2016xmw}.  Evaluating Eq.~(\ref{eq:br1}), we find $\mathcal{BR}(K_L\rightarrow\mm\gamma)=5.13(4)\times10^{-13}|f\left(z_{TM}\right)|^2$, where the dominant error is from $\mathcal{BR}(K_L\rightarrow\mm\gamma)$, preventing the measurement of these radiative corrections from this ratio.  An improved value of $\mathcal{BR}(K_L\rightarrow\mm\gamma)$ or constructing a different ratio, as we do below, can allow sensitivity to these corrections. 

The theoretical predictions for $f(z)$ are computed as a series expansion to first order in $z$ with slope $b$.  It is typically decomposed into $b=b_V+b_D$. $b_V$ arises from a weak transition from $K_L\rightarrow P$ followed by a strong-interaction vector interchange $P\rightarrow V\gamma$ and concluding with the vector meson mixing with the off-shell photon.  Here, we denote with $P$ the pseudoscalars ($\pi^0,\eta,\eta')$ and with $V$ the vector mesons ($\rho,\omega,\phi$). The second term, $b_D$, arises from the direct weak vertex $K_L\rightarrow V\gamma$ which then mixes with $\gamma+\gamma^*$ which requires modeling.  Following~\cite{DAmbrosio:1996kjn}, the predictions of $b_V$ and $b_D$ are divided into whether nonet or octet symmetry in the light mesons is assumed.  

To compute $b_V$, one integrates out the vector mesons from the $P\rightarrow V\gamma$ vertex and assuming a particular pseudoscalar symmetry, the effective Lagrangian is derived and low energy constants can be used.  $b_V^{octet}=0$ at leading order due to the cancellation between $\pi^0$ and $\eta$ in the Gell-Mann-Okubo relation~\cite{GellMann:1961ky,Okubo:1961jc}.  In the nonet realization, a nonzero contribution coming from $\eta'$ yields $b_V^{nonet}=r_V M^2_K/M^2_\rho\sim0.46$~\cite{ecker1990chiral}, where $r_V$ is a model-independent parameter depending on the couplings of each decomposed meson fields in the effective Lagrangian and are ultimately determined by experimental data.

For $b_D$, the derivation is more complicated and relies on models.  In the naive factorization model (FM)~\cite{Pich:1990mw,*Ecker:1992de,*Ecker:1993cq}, the dominant contribution to the weak vertex is assumed to be factorized current$\times$current operators which neglect the chiral structure of QCD. 
A free parameter, $k_F$, is introduced that is related to goodness of the factorized current approximation.  If this factorization was exact, $k_F=1$.  In this scheme, $b^{nonet}_D=2b^{octet}_D=1.41k_F$.  This model predicts the process $K_L\rightarrow \pi^0\gamma\gamma$ as well, and we use the unweighted average of the two most recent measurements of this process to fix $k_F= 0.55(6)$~\cite{Lai:2002kf,Abouzaid:2008xm}.

In the Bergstr\"om-Mass\'o-Singer (BMS) model~\cite{Bergstrom:1983rj,*Bergstrom:1990uh}, the direct transition is instead assumed to be dominated by a weak vector-vector interaction ($K_L\rightarrow \gamma+K^*\rightarrow\gamma+\rho,\omega,\phi\rightarrow\gamma+\gamma^*$).  BMS further assumes that no $\Delta I=\frac{1}{2}$ enhancement occurs.  This model produces a complete form factor:
\begin{align}
\label{eq:bmsff}
 f_{\gamma^*,BMS}(z)=&\frac{1}{1-\frac{M^2_K}{M^2_\rho}z}+\frac{C\,\alpha_{K^*}}{1-\frac{M^2_K}{M^2_{K^*}}z}\bigg(\frac{4}{3}-\frac{1}{1-\frac{M^2_K}{M^2_\rho}z}\nonumber\\&-\frac{1}{9}\frac{1}{1-\frac{M^2_K}{M^2_\omega}z}-\frac{2}{9}\frac{1}{1-\frac{M^2_K}{M^2_\phi}z}\bigg).
\end{align}
The two terms correspond to the vector interchange and direct transition, respectively.  Expanding this expression in powers of $z$, we find the BMS model predicts
\begin{align}
 b_{BMS}=&\frac{M^2_{K}}{M^2_{\rho}}-\frac{1}{9}C\,\alpha_{K^*}\left(9\frac{M^2_K}{M^2_\rho}+2\frac{M^2_K}{M^2_\phi}+\frac{M^2_K}{M^2_\omega}\right)\nonumber\\=&0.41205-0.509926C\,\alpha_{K^*}\nonumber\\=&b_{V,BMS}+b_{D,BMS}
\end{align}
Under the model assumptions, $-\alpha_{K^*}$ is theoretically estimated to be $\sim0.2-0.3$~\cite{Bergstrom:1983rj,*Bergstrom:1990uh}. $C=2.7(4)$ depends on a number of other mesonic decay rates~\cite{Ohl:1990qw,*AlaviHarati:2001wd}, and we used the modern values~\cite{Olive:2016xmw}. The error comes from the experimental uncertainty which is dominated by the two $K^*$ measurements. $\mathcal{BR}(K^*\rightarrow K^0\gamma)$ contributes $\Delta C\sim 13\%$ and $\Gamma_{K^*,tot}$ contributes $\Delta C\sim 4\%$ due to a disagreement between decay modes.  This choice of $C$ and $\alpha_{K^*}$ is consistent with the measured rates for $K_L\rightarrow\ell^+\ell^-\gamma$.
  
D'Ambrosio et. al. advocates the view that $b_{D,BMS}$ is one of a series of contributions to $b_D$, which should be summed together with the model-independent $b_{V}$~\cite{DAmbrosio:1996kjn}.  They construct another contribution by factorizing the vector coupling (FMV) similar to FM but first restricting the Lagrangian to left-handed currents.  For the different symmetry realizations, $b_D^{nonet}=3.14\eta\sim 0.66$ and $b_D^{octet}=2.42\eta\sim0.51$ where $\eta$ is a coefficient multiplying the naive weak coupling $G_8$ and like $k_F$ is related to the quality of the factorization assumption. We use their value of $\eta=g_8^{Wilson}/|g_8|_{K\rightarrow\pi\pi,LO}=0.21$.
Our theoretical results are compiled in Table~\ref{tab:1}.  These values disagree outside their error, and a 10\% precision measurement would be able to discriminate between them.  This is in contrast to the radiative Dalitz decays, where the theoretical values are consistent.

\begin{table}[h]
 \caption{\label{tab:1}Theoretical values of $b$ and $\mathcal{BR}(K_L\rightarrow\mm\gamma)$ for the models considered in this paper.}
 \begin{center}
 \begin{tabular}{l c c}
 \hline\hline
  $Model$&$b_{theory}$&$\mathcal{BR}_{TM}\times10^{13}$\\
  \hline
  (FM)$^{octet}$&$0.40(4)$\footnote{\label{kf}Using value of $k_F=0.55(6)$ derived from $K_L\rightarrow\pi^0\gamma\gamma$\cite{Lai:2002kf,Abouzaid:2008xm}}&5.90(9)\\
  (FM)$^{nonet}$&$1.24(6)$\footref{kf}&7.68(15)\\
  (BMS)$^{nonet}$&$0.76(9)$&6.63(20)\\
  (BMS+FMV)$^{octet}$&$0.85(10)$&6.82(22)\\
  (BMS+FMV)$^{nonet}$&$1.45(10)$&8.16(25)\\
    \hline\hline
 \end{tabular}
\end{center}
\end{table}

The BMS form factor also has been used to phenomenologically fit $K_L\rightarrow\ell^+\ell^-\gamma$ for both $\ell=e,\mu$, and $C\,\alpha_{K^*}$ is derived from the differential cross sections of these processes;  yielding $(C\alpha_{K^*})_e=-0.517(30)_{stat}(22)_{sys}$~\cite{Abouzaid:2007cm} and $(C\,\alpha_{K^*})_\mu=-0.37(7)$~\cite{AlaviHarati:2001wd}, which are each input into our prediction for $\mm$.

We also consider the D'Ambrosio-Isidori-Portol\'es (DIP) phenomenological $F_{\gamma^*\gamma^*}(z_1,z_2)$~\cite{DAmbrosio:1997eof}:  
\begin{align}
f_{\gamma^*\gamma^*,DIP}(z_1,z_2)=&\phantom{x}1+\alpha_{DIP}\left(\frac{z_1}{z_1-\frac{M^2_\rho}{M^2_K}}+\frac{z_2}{z_2-\frac{M^2_\rho}{M^2_K}}\right)\nonumber\\&\phantom{x1}+\beta_{DIP}\frac{z_1z_2}{\left(z_1-\frac{M^2_\rho}{M^2_K}\right)\left(z_2-\frac{M^2_\rho}{M^2_K}\right)}.
 \end{align}
where $z_1=z_{TM},z_2=0$ for $\mm$ production. To set $\alpha_{DIP}$, we take the values from $K_L\rightarrow e^+e^-\gamma$, $\alpha_{DIP,e}=-1.729(43)_{stat}(28)_{sys}$~\cite{Abouzaid:2007cm}, and from $K_L\rightarrow \mu^+\mu^-\gamma$, $\alpha_{DIP,\mu}=-1.54(10)$~\cite{AlaviHarati:2001wd}.  Our phenomenological results are compiled in Table~\ref{tab:2}.  Comparing the phenomenological form factors, they are indistinguishable within uncertainty in $\mm$ production.  This is perhaps unsurprising because they arise from the same underlying data, but the difference in functional forms could be discriminated by higher precision data.
\begin{table}[h]
 \caption{\label{tab:2}Values of $|f(z_{TM})|$ and $\mathcal{BR}(K_L\rightarrow\mm\gamma)$ computed using the phenomenological form factors with parameters set by either radiative $K_L$ decay to $e$ or $\mu$.}
 \begin{center}
 \begin{tabular}{l c c}
 \hline\hline  
 $Model$&$|f(z_{TM})|$&$\mathcal{BR}_{TM}\times10^{13}$\\
 \hline
 BMS$_{ee\gamma}$&1.134(6)\footnote{\label{err}The systematic and statistical errors have been summed}&6.60(10)\\
  BMS$_{\mu\mu\gamma}$&1.119(8)&6.42(11)\\
  DIP$_{ee\gamma}$&1.139(6)\footref{err}&6.66(10)\\
  DIP$_{\mu\mu\gamma}$&1.124(9)&6.48(12)\\
  \hline\hline
 \end{tabular}
\end{center}
\end{table}

Due to the small value of $z_{Ps}\approx 4M_e^2/M^2_K$, the branching ratio to positronium, $\mathcal{BR}(K_L\rightarrow(e^+e^-)\gamma)=9.31(5)\times10^{-13}$, is independent of the form factor within the error of $\mathcal{BR}(K_L\rightarrow\gamma\gamma)$ and slightly larger than $\mm$.  While this branching ratio also has not been measured, one can construct a ratio 
\begin{align}
R=&\frac{\mathcal{BR}(K_L\rightarrow\mm\gamma)}{\mathcal{BR}(K_L\rightarrow(e^+e^-)\gamma)}\nonumber\\
=&\frac{(1-z_{TM})^3\left(1-0.439\frac{\alpha}{\pi}\right)|f(z_{TM})|^2}{(1-z_{Ps})^3\left(1-\frac{52}{9}\frac{\alpha}{\pi}\right)|f(z_{Ps})|^2}\nonumber\\=&0.55767(2)\bigg|\frac{f(z_{TM})}{f(z_{Ps})}\bigg|^2,
\end{align}
which is independent of the $\mathcal{BR}(K_L\rightarrow\gamma\gamma)$ uncertainty and directly measures lepton universality without an uncertainty due to $Q^2$ binning.  By taking the largest and smallest theoretical values of $b$ to give a gross range, we predict $R=0.76(14)$.  Applying the same procedure to the phenomenological form factors yields $R=0.707(9)$.

We now focus upon the experimental situation.  Throughout, we assume a 10\% acceptance. The largest previous experimental data set that could be used to study $\mathcal{BR}(K_L\rightarrow \mm \gamma)$ is KTEV.  We estimate from the number of events reported for $\mathcal{BR}(K_L\rightarrow\ell^+\ell^-\gamma)$~\cite{AlaviHarati:2001wd,Abouzaid:2007cm} that at least 1000 times the luminosity would be required for just one $\mm$ event.  From the existing data, one might expect to place a limit on the order of $\mathcal{BR}(K_L\rightarrow \mm \gamma)\lesssim10^{-9}$.  

The KOTO experiment at J-PARC has reported $3.560(0.013)\times10^7$ $K_L$ per 2$\times10^{14}$ protons on target (POT)~\cite{Shiomi:2012zz}.  Their 2013 physics run accumulated 1.6$\times 10^{18}$ POT~\cite{Ahn:2016kja} which would correspond to 0.015 $\mm$ events.  Through their 2015 physics run, 20 times the $K_L$ decays have been recorded~\cite{Ahn:2016kja}, indicating 0.3 produced $\mm$ events and a limit of $\lesssim10^{-11}$.  Unfortunately,  the KOTO experiment is designed to detect only photons, and detecting purely photon decay products of $\mm$ would be difficult.  The J-PARC kaon beam hopes to run into the 2020s with an additional flux upgrade so a discovery is quite possible in an experiment with lepton identification.  The NA62-KLEVER proposal~\cite{Moulson:2016zsl} for a rare $K_L$ beam at CERN hopes to start by 2026 and accumulate 3$\times10^{13}$ $K_L$ over 5 years, which would also be nearly sufficient for single-event sensitivity.

A few channels are available to measure the branching ratio of true muonium: dissociated $\mu^+\mu^-$ with or without $\gamma$, decayed $e^+e^-$ with or without $\gamma$, or $\ell^\pm\gamma$ similar to SUSY searches with invisible decays~\cite{Hinchliffe:1996iu,*Allanach:2000kt}. The decay to $\pi^0\gamma$ is suppressed by $10^{-5}$ but KOTO can search for it without modification~\cite{Czarnecki:2017yde}.  

For each channel, different backgrounds matter.  The dominant backgrounds will arise from the free decays $K_L\rightarrow \ell^+\ell^-\gamma$.  We compute the branching ratio for this by integrating the differential cross section in an invariant mass bin, $M_{bin}$, around the $\mm$ peak to obtain a background estimate.  In the case of electrons, the bin is centered around the $\mm$ peak; for muon final states it is defined as $[2m_\mu,2m_\mu+M_{bin}]$.  This difference in binning reflects that the muons are above threshold.  For bin size similar to KTEV, the values are $\mathcal{BR}(K_L\rightarrow e^+e^-\gamma)_{bin}=1.2\times10^{-8}M_{bin}$, and $\mathcal{BR}(K_L\rightarrow \mu^+\mu^-\gamma)_{bin}=5.0\times10^{-9}M_{bin}$ where $M_{bin}$ is in MeV.  This large raw background ($\sim10^{5}\times$ the signal) will have to be reduced, but it has distinct features compared to true muonium decays which can be leveraged. 

The smoothness of the background differential cross section around the $\mm$ peak should allow accurate modeling from the sidebands.  Reconstruction of the $K_L$ allows the energy of the $K_L$ to be used to cut on the $\gamma$ and leptonic energies.   The two two-body decay topology suggests cuts on momenta and angular distribution would be powerful in background suppression.  As an example, for radiative Dalitz decay the angle $\theta_e$ between the electrons can be arbitrary, but from the true muonium decay $e$ will have $\theta_e\sim m_{TM}/E_{TM}\sim50^o\times\frac{\text{GeV}}{E_{K_L}}$.  This suggests the higher energy of the proposed CERN beamline would be desirable.  Additionally, vertex cuts can be made using the proper lifetime of true muonium $c\tau=0.5n^3$ mm, where $n$ is the principal quantum number.  A more rigorous study of backgrounds is planned for the future.  

\begin{acknowledgments}
HL is supported by the U.S. Department of Energy under Contract No. DE-FG02-93ER-40762.  YJ acknowledges the Deutsche Forschungsgemeinschaft for support under grant BR 2021/7-1.
\end{acknowledgments}
\bibliographystyle{apsrev4-1}
%

\end{document}